\documentclass[prd,onecolumn,nofootinbib,showkeys]{revtex4}
\usepackage{bm,amsmath,amssymb,graphicx}
\newcommand\gothfamily{\usefont{U}{ygoth}{m}{n}}
\DeclareTextFontCommand{\textgoth}{\gothfamily}

\begin{document}

\title{THE SIMPLEST ORIGIN OF THE BIG BOUNCE AND INFLATION}

\author{{\bf Nikodem Pop{\l}awski}}

\affiliation{Department of Mathematics and Physics, University of New Haven, 300 Boston Post Road, West Haven, CT 06516, USA}
\email{NPoplawski@newhaven.edu}

\noindent
{\em International Journal of Modern Physics D}\\
Vol. {\bf 27}, No. 14 (2018) 1847020\\
\copyright\,World Scientific Publishing Company
\vspace{0.4in}

\begin{abstract}
Torsion is a geometrical object, required by quantum mechanics in curved spacetime, which may naturally solve fundamental problems of general theory of relativity and cosmology.
The black-hole cosmology, resulting from torsion, could be a scenario uniting the ideas of the big bounce and inflation, which were the subject of a recent debate of renowned cosmologists.
\end{abstract}



\maketitle

I have recently read an interesting article, {\it Pop Goes the Universe}, in the January 2017 edition of Scientific American, written by Anna Ijjas, Paul Steinhardt, and Abraham Loeb (ISL) \cite{ISL1}. This article follows an article by the same authors, {\it Inflationary paradigm in trouble after Planck 2013} \cite{ISL2}. They state that cosmologists should reassess their favored inflation paradigm because it has become nonempirical science, and consider new ideas about how the Universe began, namely, the big bounce. Their statements caused a group of 33 renowned physicists, including 4 Nobel Prize in Physics laureates, to write a reply, {\it A Cosmic Controversy}, categorically disagreeing with ISL about the testability of inflation and defending the success of inflationary models \cite{reply}.

I agree with the ISL critique of the inflation paradigm, but I also agree with these 33 physicists that some models of inflation are testable. As a solution to this dispute, I propose a scenario, in which every black hole creates a new universe on the other side of its event horizon. Accordingly, our Universe may have originated from a black hole existing in another universe. This scenario considers a geometrical property of spacetime called torsion, which can generate both the big bounce and inflationary dynamics, and I published it as {\it Universe in a black hole in Einstein--Cartan gravity} in \cite{ApJ}.

The current theory of the origin of our Universe, which is based on Einstein's general theory of relativity (GR), assumes that our Universe has started more than 13 billion years ago from an extremely hot and dense state called the big bang. The big-bang cosmology successfully describes primordial nucleosynthesis (production of the lightest elements in the early Universe) and predicts the cosmic microwave background (CMB) radiation, which was emitted about 370,000 years after the big bang and which we observe coming from all directions in the sky. In order to explain why the Universe that we observe today appears at the largest scales flat (not curved) and nearly uniform all over space and in every direction, the theory of cosmic inflation has been proposed, according to which the early Universe went through an extremely accelerated (exponential) expansion by an enormous factor in volume \cite{infl}. Inflation also can predict the form of primordial density fluctuations observed in the CMB, which seed the structure formation in the Universe: stars, galaxies, and galaxy clusters.

Inflation requires that the Universe be filled with an exotic form of high-density energy that gravitationally self-repels, enhancing and speeding up the expansion of the Universe. This inflationary energy is hypothetical and we have no evidence that it exists. There have been hundreds of models regarding the origin of inflation in the last 36 years, generating different rates of inflation. The most common models attribute inflation to a hypothetical scalar field called inflaton. Even with the inflaton, inflation is not a precise theory but rather a highly flexible framework that admits many possibilities. Moreover, inflation does not tell us why the big bang happened or what created the initial volume of space that evolved into the Universe observed today.

According to the scientists analyzing the results from a Planck satellite of the European Space Agency, the new map of the CMB confirms inflation \cite{Planck2013}. ISL do not agree with this interpretation of the Planck 2013 results. These results eliminate a wide range of more complex inflationary models and favor models with a single scalar field, as reported by the Planck Collaboration. Among single-scalar-field models, Planck 2013 disfavors the simplest (power-law) inflaton models relative to models with plateau-like potentials. However, as ISL point out, plateau-like models have serious problems: they require an initially smooth Universe (the initial conditions problem), are in the class of eternally inflating models (which leads to unpredictable creation of new universes in the multiverse), and are unlikely compared to power-law inflation (have much smaller scalar-field range and amount of expansion) \cite{ISL1}. In addition, scalar-field plateau-like models require at least three parameters.

The big bang itself is also unphysical: the big-bang Universe started from a point of infinite density, called singularity.  ISL advocate for another scenario in which the Universe began with a big bounce, a transition from a contracting cosmological phase to the current expanding phase. They write \cite{ISL1}: ``Although most cosmologists assume a bang, there is currently no evidence --- zero --- to say whether the event that occurred 13.7 billion years ago was a bang or a bounce. Yet a bounce, as opposed to a bang, does not require a subsequent period of inflation to create a universe like the one we find, so bounce theories represent a dramatic shift away from the inflation paradigm.'' In bounce theories, contraction before a bounce can smooth and flatten the Universe, which is what inflation was supposed to do when it was proposed. In addition, bounce theories do not produce multiple universes which are predicted by inflationary scenarios.

The most natural theoretical-physics mechanism which generates a bounce comes from an old (the 1920s) extension of general relativity, called the Einstein--Cartan (EC) or Einstein--Cartan--Sciama--Kibble theory of gravity \cite{SK,EC}. This theory extends GR by removing its artificial symmetry constraint on the spacetime affine connection (the connection is a geometrical quantity which tells us how to do calculus in a curved space). Instead, a part of the affine connection called the torsion tensor can be different from zero and turns out to be related to the quantum-mechanical, intrinsic angular momentum of elementary particles called spin, as shown by Dennis Sciama and Tom Kibble in the 1960s \cite{SK}. Even though the spin is a quantum phenomenon, it originates, like the mass, from the Casimir invariants of the Poincar\'{e} algebra. The conservation law for the total angular momentum (orbital plus spin) of a particle in curved spacetime that admits the exchange between its orbital and intrinsic components (spin--orbit interaction) requires torsion. The field equations give a linear differential relation between the curvature and energy--momentum of matter, as in GR, and a linear algebraic relation between the torsion and spin of matter.

These two relations introduce effective corrections to the energy--momentum tensor of matter, which are significant only at extremely high densities, much larger than the density of nuclear matter, existing in black holes and near the big bang. In a May 2012 article in Inside Science, {\it Every Black Hole Contains a New Universe}, I wrote \cite{Inside}: ``In this picture, spins in particles interact with spacetime and endow it with a property called torsion. To understand torsion, imagine spacetime not as a two-dimensional canvas, but as a flexible, one-dimensional rod. Bending the rod corresponds to curving spacetime, and twisting the rod corresponds to spacetime torsion. If a rod is thin, you can bend it, but it is hard to see if it is twisted or not.''

At such high densities, torsion manifests itself as a force that counters gravity, which was discovered by Andrzej Trautman and Friedrich Hehl and their collaborators in the 1970s \cite{avert}. As in GR, very massive stars end up as black holes: regions of space from which nothing, not even light, can escape. Gravitational attraction due to curvature initially overcomes repulsion due to torsion and matter in a black hole collapses, but eventually the coupling between torsion and spin (acting like gravitational repulsion) becomes very strong and prevents the matter from compressing indefinitely to a singularity. The matter instead reaches a state of finite, extremely large density, stops collapsing, undergoes a bounce like a compressed spring, and starts rapidly expanding. Extremely strong gravitational forces near this state cause an intense, quantum particle production, increasing the mass inside a black hole by many orders of magnitude and strengthening gravitational repulsion that powers the bounce. The rapid recoil after such a big bounce could be what has led to our expanding Universe. It also explains why the present large-scale Universe appears at flat and nearly uniform all over space and in every direction, without needing scalar-field inflation, which I showed in: {\it Cosmology with torsion: An alternative to cosmic inflation} \cite{cosmology}.

The energy of matter at the big bounce is an order of magnitude higher than the Planck energy. Recent observations of high-energy photons from gamma-ray bursts, however, indicate that spacetime may behave classically even at scales above the Planck energy. The classical spin-torsion mechanism of the bounce may thus be justified. Furthermore, the EC theory passes all tests of GR because both theories give significantly different predictions only at extremely high densities that exist in black holes or in the very early Universe.

Torsion in the EC gravity therefore provides a theoretical explanation of a scenario (suggested also by Igor Novikov, Lee Smolin, and Stephen Hawking \cite{BH}), according to which every black hole produces a new, baby universe inside and becomes an Einstein--Rosen bridge (wormhole) that connects this universe to the parent universe in which the black hole exists: {\it Radial motion into an Einstein--Rosen bridge} \cite{bridge}. In the new universe, the parent universe appears as the other side of the only white hole, a region of space that cannot be entered from the outside and which can be thought of as the time reverse of a black hole. Accordingly, our own Universe could have originated from the interior of a black hole existing in another universe (our Universe as the interior of a black hole was proposed by Raj Pathria \cite{Pat}). The motion of matter through the black hole's boundary called an event horizon can only happen in one direction, providing a past--future asymmetry at the horizon and thus everywhere in the baby universe. The arrow of time in such a universe would therefore be inherited, through torsion, from the parent universe \cite{cosmology}.

The new universe in a black hole is closed: finite and without boundaries (with the exception of the white hole, which connects it to the parent universe). It can be thought of as a three-dimensional analogue of the two-dimensional surface of a sphere: if you continue to go in any direction, you eventually would come back from the opposite direction. The formation and evolution of such a universe is not visible for external observers in the parent universe, for whom the event horizon's formation and all subsequent processes would occur after an infinite amount of time had elapsed (because of the time dilation by gravity). A baby universe is thus a separate branch of spacetime with its own timeline. ISL write \cite{ISL1}: ``bouncing theories have an important advantage compared with inflation: they do not produce a multimess.'' The black-hole cosmology presented here produces other universes but only beyond the event horizons of black holes, thus the resulting multiverse is organized.

In a recent article, {\it Non-parametric reconstruction of an inflaton potential from Einstein--Cartan--Sciama--Kibble gravity with particle production}, written with Shantanu Desai \cite{SD}, we analyzed numerically the dynamics of the early Universe based on the EC theory of gravity with quantum particle production and proposed in my ApJ 2016 article \cite{ApJ}. The scale factor of the Universe $a$ and its temperature $T$ satisfy the Friedmann equations:
\begin{eqnarray}
& & \frac{{\dot{a}}^2}{c^2}+1=\frac{1}{3}\kappa\tilde{\epsilon}a^2=\frac{1}{3}\kappa(h_\star T^4-\alpha h_{n\textrm{f}}^2 T^6)a^2, \label{dynamics1} \\
& & \frac{\dot{a}}{a}+\frac{\dot{T}}{T}=\frac{cK}{3h_{n1}T^3},
\label{dynamics2}
\end{eqnarray}
where dot denotes the time derivative, $\tilde{\epsilon}$ is the effective energy density (the positive term proportional to $T^4$ comes from relativistic matter and the negative term proportional to $T^6$ comes from the spin-torsion gravitational repulsion), $\kappa=8\pi G/c^4$, $h$-constants depend on the number of the thermal degrees of freedom (the number of elementary particle species), and $K=\beta(\kappa\tilde{\epsilon})^2$ is the particle production rate. This dynamics contains only one unknown parameter, the particle production coefficient $\beta$. We found that for different values of this coefficient, the universe can have different numbers of cycles, each of which begins with a nonsingular bounce followed by expansion to a crunch, followed by contraction to the next bounce. Each new cycle, compared to the previous cycle, lasts longer and represents the universe with a larger amount of matter. The last bounce: the big bounce, is what we refer to as the big bang. From the obtained time dependence of the scale factor (size) of the universe, we reconstructed a scalar field potential which would give the same dynamics of the early universe. We did it because the CMB quantities measured by the Planck satellite can be directly calculated for scalar-field inflation models.

\begin{figure}
\centering
\includegraphics[width=0.5\textwidth]{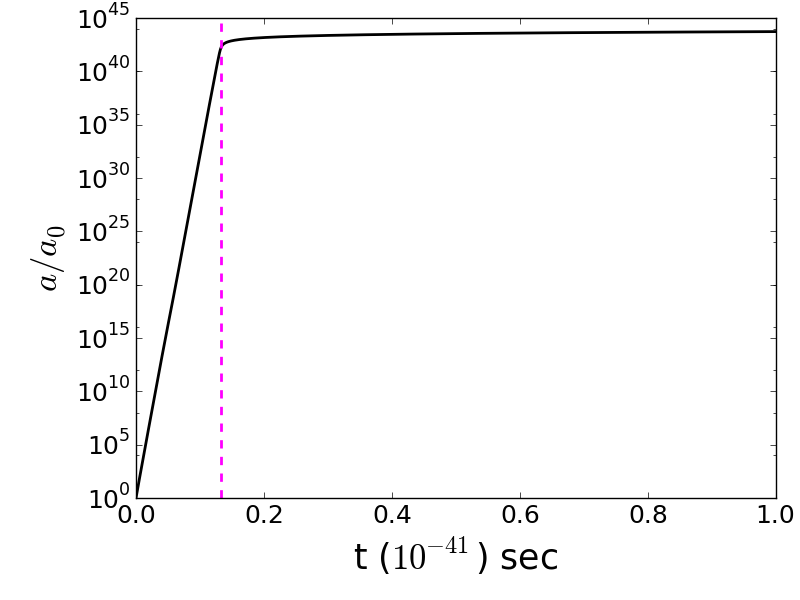}
\caption{The ratio of the scale factor $a(t)$ to its initial value (at the big bounce) as a function of time for a particlular value of the particle production coefficient $\beta$. The dashed magenta line represents the transition from an accelerating (torsion-dominated) phase to a decelerating (radiation-dominated) phase. We obtain about 60 $e$-folds.}
\label{scale}
\end{figure}

\begin{figure}
\centering
\includegraphics[width=0.5\textwidth]{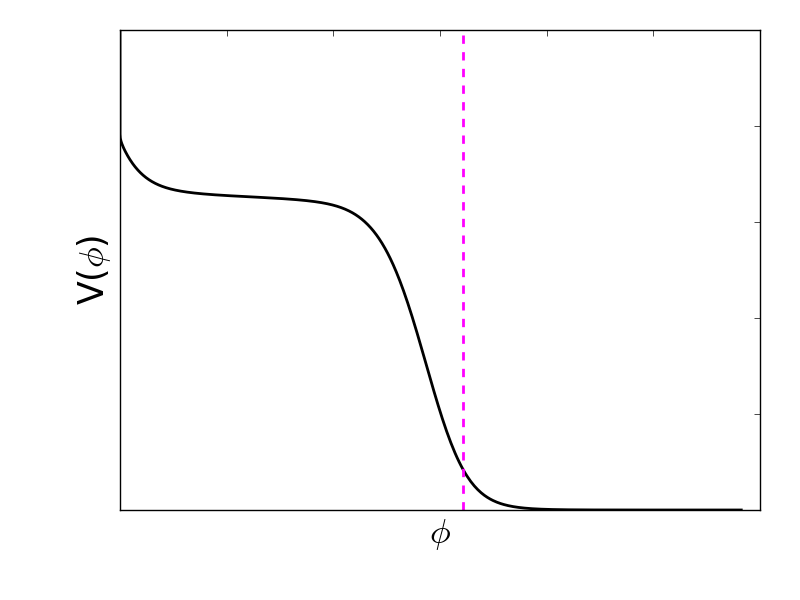}
\caption{The scalar-field potential $V(\phi)$ giving the same dynamics as Eqs. (\ref{dynamics1}) and (\ref{dynamics2}). The vertical line indicates the end of inflation when the Universe transitions from acceleration to deceleration. The shape of the potential is plateau-like.}
\label{potential}
\end{figure}

For a particular range of the particle production coefficient, we obtained one bounce (the big bounce) followed by a nearly exponential expansion of the Universe, which lasted for an extremely short amount of time and which smoothed out and flattened the Universe to the observed values, as shown in Fig. \ref{scale} \cite{SD}. We found that the reconstructed potential is of a plateau-like shape, as shown in Fig. \ref{potential} \cite{SD}, which is supported by the Planck 2013 data \cite{Planck2013}. From this potential, we calculated the CMB quantities measured by the Planck satellite, and found that they do not significantly depend on the scale factor at the big bounce. Our predictions for these quantities are consistent with the Planck 2015 observations \cite{Planck2015}. This scenario can thus produce big bounce and plateau-like inflation without a scalar field.

Unlike plateau-like models based on a scalar field, our scenario escapes the problems noted by ISL: it does not rely on a hypothetical scalar field, does not depend on the initial conditions of the Universe, avoids eternal inflation, and is simpler --- has only one parameter (which should be predicted by a future quantum theory of gravity). And most importantly, it eliminates the initial singularity problem, which is not addressed by scalar-field inflation \cite{ApJ}. This scenario is motivated by including quantum-mechanical spin in a theory of gravity, and the nonsingular big bounce and inflation are derived, not imposed. Inflation models assume that the inflaton starts inflation and then somehow decays into ordinary matter. The EC theory gives a simpler dynamics: spin-generated torsion produces the bounce, quantum particle production starts inflation (which was also suggested in \cite{Sing}), weakening of torsion ends inflation (without reheating), and weakening of curvature ends particle production, after which the Universe expands through standard radiation-dominated and matter-dominated eras.

In addition, torsion modifies the Dirac equation that describes the quantum-mechanical behavior of fermions (quarks and leptons) that form ordinary matter. These particles must be spatially extended, which may solve the ultraviolet-divergence problems in quantum field theory arising from treating them as points. The spatial extension of fermions arising from torsion is at the level of the Cartan length, which for an electron is about 100 million Planck lengths and can be tested in the future: {\it Nonsingular Dirac particles in spacetime with torsion} \cite{nonsingular}. This modification may also be responsible for the observed imbalance of matter and antimatter in the Universe and could relate the apparently missing antimatter to dark matter, a mysterious form of matter that does not interact electromagnetically and accounts for a majority of the matter in the Universe: {\it Matter--antimatter asymmetry and dark matter from torsion} \cite{matter}. Finally, torsion may be the source of dark energy, a mysterious form of energy that permeates all of space and increases the present rate of expansion of the Universe, allowing it to grow infinitely large and to last infinitely long: {\it Affine theory of gravitation} \cite{affine}.

ISL wrote \cite{ISL1}: ``The fact that our leading ideas have not worked out is a historic opportunity for a theoretical breakthrough. Instead of closing the book on the early universe, we should recognize that cosmology is wide open''. The 33 physicists wrote in their reply \cite{reply}: ``Like any scientific theory, inflation need not address all conceivable questions. Inflationary models, like all scientific theories, rest on a set of assumptions, and to understand those assumptions we might need to appeal to some deeper theory. This, however, does not undermine the success of inflationary models.''

Here is my response: The EC theory of gravity may be that deeper theory which has been waiting for its breakthrough. Spacetime torsion is a geometrical phenomenon, required by quantum mechanics, which may naturally solve fundamental problems of GR and cosmology. It provides the simplest and most natural mechanism for a bounce that started the expansion of our Universe. With quantum particle production, it also provides the simplest model of inflation that does not need hypothetical fields, lasts for a finite time, and is consistent with the Planck data. The black-hole cosmology could be a scenario uniting the viewpoints of the two groups of physicists debating in Scientific American.

\section*{Acknowledgment}

This work was funded by the University Research Scholar program at the University of New Haven.


\begin{thebibliography}{}
\bibitem{ISL1} A. Ijjas, P. J. Steinhardt, and A. Loeb, {\it Pop Goes the Universe}, Scientific American, January 2017.
\bibitem{ISL2} A. Ijjas, P. J. Steinhardt, and A. Loeb, Phys. Lett. B {\bf 723}, 261 (2013).
\bibitem{reply} A. H. Guth {\it et al.}, {\it A Cosmic Controversy}, Scientific American, May 2017.
\bibitem{ApJ} N. Pop{\l}awski, Astrophys. J. {\bf 832}, 96 (2016).
\bibitem{infl} D. Kazanas, Astrophys. J. {\bf 241}, L59 (1980); A. H. Guth, Phys. Rev. D {\bf 23}, 347 (1981); A. Linde, Phys. Lett. B {\bf 108}, 389 (1982).
\bibitem{Planck2013} P. A. R. Ade {\it et al.}, Astron. Astrophys. {\bf 571}, A16 (2014).
\bibitem{SK} T. W. B. Kibble, J. Math. Phys. {\bf 2}, 212 (1961); D. W. Sciama, in {\em Recent Developments in General Relativity}, 415 (Pergamon, 1962); Rev. Mod. Phys. {\bf 36}, 463 (1964); Rev. Mod. Phys. {\bf 36}, 1103 (1964).
\bibitem{EC} E. A. Lord, {\em Tensors, Relativity and Cosmology} (McGraw-Hill, 1976); F. W. Hehl, P. von der Heyde, G. D. Kerlick, and J. M. Nester, Rev. Mod. Phys. {\bf 48}, 393 (1976); V. de Sabbata and M. Gasperini, {\it Introduction to Gravitation} (World Scientific, 1985); M. Gasperini, Phys. Rev. Lett. {\bf 56}, 2873 (1986); V. de Sabbata and C. Sivaram, {\it Spin and Torsion in Gravitation} (World Scientific, 1994); N. J. Pop{\l}awski, arXiv:0911.0334.
\bibitem{Inside} N. Pop{\l}awski, {\it Every Black Hole Contains a New Universe}, Inside Science, May 2012.
\bibitem{avert} W. Kopczy\'{n}ski, Phys. Lett. A {\bf 39}, 219 (1972); Phys. Lett. A {\bf 43}, 63 (1973); A. Trautman, Nature (Phys. Sci.) {\bf 242}, 7 (1973); F. W. Hehl, P. von der Heyde, and G. D. Kerlick, Phys. Rev. D {\bf 10}, 1066 (1974); B. Kuchowicz, Gen. Relativ. Gravit. {\bf 9}, 511 (1978).
\bibitem{cosmology}  N. J. Pop{\l}awski, Phys. Lett. B {\bf 694}, 181 (2010); Phys. Lett. B {\bf 701}, 672 (2011).
\bibitem{BH} I. D. Novikov, J. Exp. Theor. Phys. Lett. {\bf 3}, 142 (1966); L. Smolin, Class. Quantum Grav. {\bf 9}, 173 (1992); S. Hawking, {\it Black Holes and Baby Universes and other Essays} (Bantam Dell, 1993).
\bibitem{bridge} N. J. Pop{\l}awski, Phys. Lett. B {\bf 687}, 110 (2010).
\bibitem{Pat} R. K. Pathria, Nature {\bf 240}, 298 (1972).
\bibitem{SD} S. Desai and N. J. Pop{\l}awski, Phys. Lett. B {\bf 755}, 183 (2016).
\bibitem{Planck2015} P. A. R. Ade {\it et al.}, Astron. Astrophys. {\bf 594}, A20 (2016).
\bibitem{Sing} J. A. S. Lima and D. Singleton, Phys. Lett. B {\bf 762}, 506 (2016).
\bibitem{nonsingular} N. J. Pop{\l}awski, Phys. Lett. B {\bf 690}, 73 (2010); Phys. Lett. B {\bf 727}, 575 (2013).
\bibitem{matter} N. J. Pop{\l}awski, Phys. Rev. D {\bf 83}, 084033 (2011).
\bibitem{affine} N. Pop{\l}awski, Gen. Relativ. Gravit. {\bf 46}, 1625 (2014).
\end{thebibliography}
\end{document}